\documentstyle[aps,pra,preprint,eqsecnum]{revtex}

\def\S{{\cal S}}
\def\A{{\cal A}}
\def\H{{\cal H}}

\newcommand{\ho}{\scriptscriptstyle}

\begin{document}

\tightenlines

\title{Entanglement of Formation of an Arbitrary State of Two Rebits%
\thanks{The author ordering on this paper is dictated by CAF's adherence
to alphabetical ordering.  CMC, operating under equally valid, but
less strongly held principles, would have preferred, in this case,
inverse alphabetical ordering.}
}

\author{
Carlton M.~Caves,$^{(1)}$\thanks{E-mail: caves@tangelo.phys.unm.edu}
Christopher~A. Fuchs,$^{(2)}$\thanks{Permanent Address: Bell Labs,
Lucent Technologies, Murray Hill, New Jersey 07974.} and
Pranaw~Rungta$^{(1)}$\thanks{E-mail: pranaw@unm.edu} }

\address{
$^{(1)}$Center for Advanced Studies, Department of Physics and
Astronomy,\\
University of New Mexico, Albuquerque, NM~87131--1156, USA\\
$^{(2)}$Los Alamos National Laboratory, MS--B285, Los Alamos, NM 87545,
USA
}

\date{2000 September 6}

\maketitle

\begin{abstract}
We consider entanglement for quantum states defined in vector spaces
over the real numbers.  Such real entanglement is different from
entanglement in standard quantum mechanics over the complex
numbers.  The differences provide insight into the nature of
entanglement in standard quantum theory.  Wootters [Phys.\ Rev.\
Lett.\ {\bf 80}, 2245 (1998)] has given an explicit formula for the
entanglement of formation of two qubits in terms of what he calls
the {\it concurrence\/} of the joint density operator. We give a
contrasting formula for the entanglement of formation of an arbitrary
state of two ``rebits,'' a rebit being a system whose Hilbert space
is a 2-dimensional real vector space.
\end{abstract}

\section{Introduction}

One of the key distinguishing features of quantum mechanics, not
found in classical physics, is the possibility of entanglement
between subsystems.  The significance of this phenomenon is now
unquestioned, as it lies at the core of several of the most important
achievements of quantum information science~\cite{Lo1998}, such as
quantum teleportation~\cite{Bennett1993} and quantum error
correction~\cite{Shor1995,Steane1996}.  Yet can we say that we understand 
the distinction between those physical theories with entanglement and 
those without?  It is difficult to claim such understanding, as our 
most well studied foil theory to date---namely, classical physics---is 
completely devoid of the phenomenon.  This paper, in a small way, 
contributes to filling that gap by studying entanglement in real
vector spaces, where though there is entanglement, it is different 
from the entanglement in the complex vector spaces of standard
quantum theory.

Our specific objective here is to analyze entanglement for states of two
``rebits,'' a rebit being a two-state system whose Hilbert space is
defined over the field of real numbers.  In particular, we give an
explicit formula for the entanglement of formation of an arbitrary
(mixed) state of two rebits.  This formula is similar in some
structural ways, but not identical to Wootters's formula for the 
entanglement of formation of two qubits~\cite{Wootters1998}. The reason 
for our considering real vector spaces is just the reason described 
above: they provide an easy, well defined foil with which to compare 
standard quantum theory; indeed, they have already been fruitful in that
regard~\cite{Wootters90}. A quantum theory that uses real vector
spaces is similar to, but not identical to the standard theory
\cite{Adler95,Stueckelberg60}. In general, we follow the philosophy of
Weinberg in Ref.~\cite{Weinberg89} in the hope that this exercise will
help identify those aspects of entanglement that are unique, those
that are accidental, and those that are necessary to the standard
theory.

To state our main result, we must build some concepts and notation.
Consider a bipartite composite system, made up of subsystems $A$ and
$B$. A density operator $\rho^{\ho AB}$ of the composite system,
pure or mixed, is said to be {\it separable\/} if it can be thought
of as arising from an ensemble of product states, i.e.,
\begin{equation}
\rho^{\ho AB}=\sum_jp_j\rho_j^{\ho A}\otimes\rho_j^{\ho B}\;.
\end{equation}
A separable pure state is itself a product state.  The reason this
definition is interesting is because a separable state can be created
by procedures that are local to each subsystem, whereas a
nonseparable state cannot be created by any local means.

Taking the matrix transpose of any density operator relative to some
orthonormal basis---this is the same as taking the complex conjugate
in that basis---yields another density operator, i.e., another
positive semi-definite operator with unit trace.  Similarly, if a
state of a bipartite system is separable, taking the {\it partial
transpose\/} on system $B$ in any basis also yields another
density operator. If, however, taking the partial transpose leads to
an operator that is not positive semi-definite, one can be sure that
the original state was an entangled state. This is the partial
transpose condition of Peres~\cite{Peres1996}. Unfortunately, for
subsystems $A$ and $B$ of arbitrary Hilbert-space dimensions, the
converse of the Peres condition is not true---the partial transpose
of an entangled state can give another positive
semi-definite operator. Thus the Peres condition does not provide a
general criterion for testing entanglement. For $2\times2$ systems
(two qubits) or $2\times3$ systems (a qubit and a qutrit), however,
the Peres condition does provide a criterion for entanglement: a
state of such a composite system is entangled if and only if its
partial transpose is a nonpositive operator, i.e., has at least one
negative eigenvalue~\cite{Horodecki1996}.

The chief resource-based measures of entanglement are the 
{\it entanglement of formation\/} and the {\it distillable entanglement\/}
\cite{Bennett1996,Cbennett1996}.  For two $d$-dimensional systems,
the pure state
\begin{equation}
|\Psi\rangle={1\over\sqrt d}\sum_{j=1}^d|e_j^{\ho
A}\rangle\otimes|e_j^{\ho B}\rangle\;,
\end{equation}
where $|e_j^{\ho A}\rangle$ and $|e_j^{\ho B}\rangle$ are orthonormal
bases for the two subsystems, is maximally entangled in the sense
that it can be used to teleport the state of another $d$-dimensional
system. The degree of entanglement of such a maximally entangled
state is $\log_2 d$, the marginal entropy of each subsystem. Suppose
that starting with $m$ such maximally entangled states, one has a
procedure, involving only local operations and classical
communication between subsystems, for creating $n$ copies of an
arbitrary state $\rho^{\ho AB}$. The entanglement of formation,
$E(\rho^{\ho AB})$, is defined to be $\log_2 d$ times the asymptotic
ratio $m/n$ for an optimal procedure, i.e., one that has the
smallest such ratio.  Similarly, suppose that starting with $n$
copies of the state $\rho^{\ho AB}$, one has a procedure, again
involving only local operations and classical communication, for
distilling $m$ maximally entangled states. The distillable
entanglement, $D(\rho^{\ho AB})$, is defined to be $\log_2 d$ times
the asymptotic ratio $m/n$ for an optimal procedure, i.e., one that
has the largest such ratio.

A separable state has no entanglement of either sort, whereas a
nonseparable state necessarily has a nonzero entanglement of
formation. For pure states, the formation process is reversible, so
the entanglement of formation and the distillable entanglement are
the same.  For mixed states, however, the distillable entanglement is
generally less than the entanglement of formation, reflecting the
irreversibility of the formation process.  Interestingly, a state
with a positive partial transpose has no distillable entanglement
\cite{Horodecki1998}.  For $2\times2$ and $2\times3$ systems, all
entangled states have a nonpositive partial transpose, as noted
above, and they also have a nonzero distillable entanglement.  For
$3\times3$ and higher-dimensional systems, however, there are
entangled states that have positive partial transpose; though these
states have nonzero entanglement of formation, they have no
distillable entanglement.  This kind of entanglement, from which no
pure-state entanglement can be distilled, is called {\it bound
entanglement}~\cite{Horodecki1998}.

The entanglement of formation of a pure state $|\Psi\rangle$ of a
bipartite system is given by the entropy of the marginal density
operators, $\rho^{\ho A}$ and $\rho^{\ho B}$:
\begin{equation}
E(\Psi) =-{\rm tr}(\rho^{\ho A}\log_2 \rho^{\ho A}) =-{\rm
tr}(\rho^{\ho B}\log_2 \rho^{\ho B}) \;.
\end{equation}
For a bipartite mixed state the entanglement of formation is more
complicated. A mixed state $\rho^{\ho AB}$ has an ensemble
decomposition in terms of pure states $|\Psi_j\rangle$, with
probabilities $p_j$, if it can be written as
\begin{equation}
\rho^{\ho AB}=\sum_j p_j|\Psi_j\rangle\langle\Psi_j| \;.
\end{equation}
Modulo a presently unanswered question about the super-additivity of
the entanglement of formation \cite{Hayden00}, the entanglement of
formation of $\rho^{\ho AB}$ is given by the minimum average
entanglement of formation of the pure states in an ensemble
\cite{Cbennett1996},
\begin{equation}
\label{eq:min} E(\rho^{\ho AB})=
\min_{\{p_j, |\Psi_j\rangle\}}\sum_jp_jE(\Psi_j) \;,
\end{equation}
where the minimum is taken over all possible ensemble decompositions.
For two qubits, Wootters has given an explicit formula for the
entanglement of formation in terms of what he calls the {\it concurrence\/} 
of the joint density operator \cite{Wootters1998,Hill1997} [see 
Eq.~(\ref{eq:compconcurrence}) for Wootters's concurrence expression].  
There are no known explicit formulae for distillable entanglement, even for 
$2\times2$ systems.  We now have all the facts about entanglement we need 
for posing the questions addressed in this paper.

Let us turn now to issues relevant for distinguishing the theory of
real quantum mechanics from standard complex quantum mechanics. The
vector space of operators on a $d$-dimensional vector space (real or
complex) is the direct sum of two natural subspaces, the space $\S$
of real, symmetric matrices, which has dimension ${1\over2}d(d+1)$,
and the space $\A$ of real, antisymmetric matrices, which has
dimension ${1\over2}d(d-1)$.  If the vector space is over the real
numbers, all the quantum states and observables lie in the symmetric
subspace.  For a complex vector space, the states and observables
are described by Hermitian operators; the (real) vector space $\H$
of Hermitian operators takes advantage of both the symmetric and
antisymmetric subspaces, it being the direct sum $\H=\S\oplus i\A$.

This is of significance for entanglement for the following reason.
Suppose one combines two systems, with dimensions $d_{\ho A}$ and
$d_{\ho B}$, to make a composite system with dimension $d_{\ho
A}d_{\ho B}$. In the complex case, the composite vector space of
Hermitian operators, of dimension $d_{\ho A}^2d_{\ho B}^2$, is the
tensor product of the corresponding spaces for $A$ and $B$, i.e.,
$\H_{AB}=\H_A\otimes\H_B$. In contrast, the composite space of
symmetric matrices, of dimension ${1\over2}d_{\ho A}d_{\ho B}(d_{\ho
A}d_{\ho B}+1)$, is not just the tensor product of the symmetric
spaces of $A$ and $B$, but rather is given by the direct sum
\begin{equation}
\S_{AB}=(\S_A\otimes\S_B)\oplus(\A_A\otimes\A_B) \;.
\end{equation}
Joint states that have a component in the doubly antisymmetric space,
$\A_A\otimes\A_B$, are necessarily entangled relative to the real vector
space, since product states cannot have a component in
$\A_A\otimes\A_B$.

Since any operation in the real vector space can be used in the
associated complex vector space, an optimal real procedure, either
for entanglement of formation $E$ or for distillable entanglement
$D$, is never better than an optimal complex procedure.  The
consequence is that for a joint state $\rho^{\ho AB}$ in the real
vector space, the real ($R$) and complex ($C$) entanglement measures
satisfy the following inequalities:
\begin{equation}
E_R(\rho^{\ho AB})\ge E_C(\rho^{\ho AB})\ge D_C(\rho^{\ho AB})\ge
D_R(\rho^{\ho AB})\;.
\label{eq:ineq}
\end{equation}

These ideas are made concrete by considering rebits. The
three-dimensional space of real, symmetric matrices is spanned by
the unit matrix $I$ and two Pauli matrices, $\sigma_x$ and
$\sigma_z$, whereas the one-dimensional space of real, antisymmetric
matrices is spanned by $i\sigma_y$. For two rebits, the
nine-dimensional space $\S_A\otimes\S_B$ is spanned by the matrices
\begin{equation}
\matrix{
I\otimes I\hfill & \sigma_x\otimes I\hfill & \sigma_z\otimes I\hfill \cr
I\otimes\sigma_x \quad\hfill & \sigma_x\otimes \sigma_x\quad\hfill &
\sigma_z\otimes\sigma_x\quad\hfill \cr
I\otimes\sigma_z\hfill & \sigma_x\otimes \sigma_z\hfill & 
\sigma_z\otimes \sigma_z\hfill}
,
\end{equation}
but the entire symmetric subspace $\S_{AB}$ includes one additional
basis
matrix,
\begin{equation}
i\sigma_y\otimes i\sigma_y=-\sigma_y\otimes\sigma_y
\;.
\end{equation}
Any state of the composite system that contains a
$\sigma_y\otimes\sigma_y$ component is necessarily entangled
relative to the real vector space, simply because tensor products of
real states can never sum up to give a $\sigma_y\otimes\sigma_y$
component.

In Sec.~\ref{sec:tworebit} we show that the entanglement of a
two-rebit state is determined entirely by the
$\sigma_y\otimes\sigma_y$ component of the state: a state $\rho^{\ho
AB}$ is separable if and only if ${\rm tr}(\rho^{\ho
AB}\sigma_y\otimes\sigma_y)=0$.  Furthermore, there is a {\it concurrence}, 
defined by
\begin{equation}
C(\rho^{\ho AB})\equiv|{\rm tr}(\rho^{\ho AB}\sigma_y\otimes\sigma_y)| \;,
\label{eq:realc}
\end{equation}
which gives the two-rebit entanglement of formation in the same way
that Wootters's concurrence gives the two-qubit entanglement of
formation.  In a concluding section (Sec.~\ref{sec:disc}), we
discuss implications of our main result.

For the present, however, it should be noted that the expression in
Eq.~(\ref{eq:realc}) {\it does not\/} correspond to Wootters'
concurrence formula for qubits simply restricted to real density
operators. This point is nicely illustrated by the real density
operator
\begin{equation}
\rho^{\ho AB}
=\frac{1}{4}\left(I{\otimes}I+\sigma_y\otimes\sigma_y\right)\;.
\end{equation}
This state is a separable state relative to complex vector space, as
can be checked by the partial transpose condition or by noting that
it can be derived from an ensemble of two product states:
\begin{equation}
\rho^{\ho AB}={1\over2}\left(
\textstyle{{1\over2}}(I+\sigma_y)\otimes\textstyle{{1\over2}}(I+\sigma_y
)+
\textstyle{{1\over2}}(I-\sigma_y)\otimes\textstyle{{1\over2}}(I-\sigma_y
)
\right) \;.
\label{Hooter}
\end{equation}
In this (eigen)decomposition, the density operator $\rho^{\ho AB}$ looks
like it comes from the mixture of two spin states: both particles
pointing in the $+y$ direction or both pointing in the $-y$ direction.

In contrast to this, the ensemble decomposition in Eq.~(\ref{Hooter})
is not allowed relative to real vector space quantum mechanics. The
real concurrence, Eq.~(\ref{eq:realc}), of this state is in fact
$C=1$, which means that the state is maximally entangled relative to
real vector space.  Perhaps more interestingly, this state is also a
bound entangled state relative to the reals.  This follows because it
is separable relative to the complex numbers, i.e., it has no complex
entanglement of formation, and hence, by the chain of inequalities
in Eq.~(\ref{eq:ineq}), it has no real distillable entanglement.

\section{Two-Rebit Entanglement of Formation}
\label{sec:tworebit}

In this section we first review Wootters's spin-flip operation and
how it leads to the real concurrence~(\ref{eq:realc}), and we then
prove our main result, an explicit formula for the real entanglement
of formation of an arbitrary state of two rebits.

The spin-flip operation for a single qubit is the anti-unitary
operator ${\cal S}=i\sigma_y{\cal C}$, where ${\cal C}$ denotes
complex conjugation in the eigenbasis of $\sigma_z$.  For a quantum
state $\rho$ of a bipartite system---we now drop the superscript
$AB$ to reduce clutter in the notation---the spin-flipped density
operator, distinguished by a tilde, is
\begin{equation}
\tilde{\rho}=(\sigma_y\otimes\sigma_y){\cal
C}(\rho)(\sigma_y\otimes\sigma_y)\;.
\end{equation}
The concurrence of a bipartite pure state $|\Psi\rangle$ is defined
to be
\begin{equation}
C(\Psi)\equiv|\langle\Psi|{\cal S}|\Psi\rangle|\;.
\end{equation}
Defining the concurrence of a mixed state $\rho$ by
\begin{equation}
C(\rho)\equiv\min_{\{p_j, |\Psi_j\rangle\}}\sum_jp_j C(\Psi_j) \;,
\label{Norbert}
\end{equation}
where the minimum is taken over all possible ensemble decompositions
of $\rho$, Wootters \cite{Wootters1998} showed that $C(\rho)$ is
given by the explicit expression
\begin{equation}
C_W(\rho) = \max\left(0,\sqrt{\lambda_1}-\sqrt{\lambda_2}-
\sqrt{\lambda_3}-\sqrt{\lambda_4}\,\right)\;,
\label{eq:compconcurrence}
\end{equation}
where the $\lambda_i$ are the (positive) eigenvalues of the operator
$\rho\tilde\rho$ (or of the operator $\sqrt\rho\tilde\rho\sqrt\rho$)
listed in order of decreasing magnitude.

For rebit states, the complex conjugation has no effect, and the
spin flip simplifies to $i\sigma_y$, i.e., a $180^\circ$ rotation
about the $y$ axis.  The differences between concurrence and
entanglement of formation for rebits and qubits can ultimately be
traced to the fact that the spin flip for qubits is an anti-linear,
as opposed to a linear operator. Hence, for a single-rebit pure state
$|\psi\rangle$, the spin-flipped state is
\begin{equation}
|\tilde{\psi}\rangle\equiv i\sigma_y|\psi\rangle
\;.
\end{equation}
For a joint pure state $|\Psi\rangle$ of two rebits, we again
define the {\it concurrence\/} to be the overlap between
$|\Psi\rangle$ and the spin-flipped state
$|\tilde\Psi\rangle=-\sigma_y\otimes\sigma_y|\Psi\rangle$, i.e,
\begin{equation}
C(\Psi)\equiv|\langle\Psi|\tilde\Psi\rangle|=
|\langle\Psi|\sigma_y\otimes\sigma_y|\Psi\rangle|
\;,
\end{equation}
and also define the concurrence of a mixed state according to
Eq.~(\ref{Norbert}).

The joint pure state $|\Psi\rangle$ can be written in terms of a
Schmidt decomposition,
\begin{equation}
|\Psi\rangle=
a_1|e_1\rangle\otimes|f_1\rangle+a_2|e_2\rangle\otimes|f_2\rangle\;,
\end{equation}
where $|e_j\rangle$ and $|f_j\rangle$ are the (real) orthonormal
eigenvectors of the marginal density operators for systems $A$ and
$B$, and $a_1$ and $a_2$ are the (positive) square roots of the
corresponding eigenvalues.  It is easy to verify that the
concurrence of $|\Psi\rangle$ is $C(\Psi)=2a_1a_2$.  Thus, as noted
by Wootters, the concurrence itself can serve as a measurement of
entanglement, varying smoothly from $0$ for product pure states to
$1$ for maximally entangled pure states.  Indeed, the entanglement
of formation of the pure state $|\Psi\rangle$ can be expressed in the
form
\begin{equation}
\label{eq:convex}
E(\Psi)=
H(a_1^2)=
H\!\left(1+\sqrt{1-C^2}\over2\right)
\equiv{\cal E}[C(\Psi)]\;,
\end{equation}
where
\begin{equation}
H(x)\equiv-x\log_{2} x-(1-x)\log_2 (1-x)
\end{equation}
is the binary Shannon entropy.  The function ${\cal E}(C)$ is
monotonically increasing and convex on the interval $0\le C\le1$.

Before proceeding to the problem at hand, let us delineate a few
facts about ensemble decompositions of density operators in a real 
Hilbert space quantum mechanics. Consider a mixed state $\rho$, defined
on a real Hilbert space of dimension $d$, whose eigendecomposition is
\begin{equation}
\rho= \sum_{j=1}^n \mu_j|\hat e_j\rangle\langle\hat e_j|
=\sum_{j=1}^n |e_j\rangle\langle e_j|\;.
\label{eq:eigendecomp}
\end{equation}
Here the vectors $|\hat e_j\rangle$ are the orthonormal eigenvectors
of $\rho$, with corresponding eigenvalues $\mu_j$, and the vectors
$|e_j\rangle\equiv\sqrt{\mu_j}|\hat e_j\rangle=\rho^{1/2}|\hat
e_j\rangle$ are subnormalized eigenvectors.  In
Eq.~(\ref{eq:eigendecomp}) the sum includes only eigenvectors with
nonzero eigenvalues, $n\le d$ thus being the rank of $\rho$.  We can
restrict attention to the support of $\rho$, the subspace of
dimension $n$ spanned by eigenvectors with nonzero eigenvalue.  On
this subspace $\rho$ has a well defined inverse.

Consider now an arbitrary pure-state ensemble decomposition of $\rho$,
\begin{equation}
\rho=
\sum_{j=1}^m p_j|\hat w_j\rangle\langle \hat w_j|=
\sum_{j=1}^m |w_j\rangle\langle w_j|
\;.
\label{eq:psdecomp}
\end{equation}
The ensemble includes $m\ge n$ normalized vectors $|\hat w_j\rangle$, with
probabilities $p_j$.  The vectors $|w_j\rangle\equiv\sqrt{p_j}|\hat
w_j\rangle$ are subnormalized, their lengths giving the
probabilities.  The real version of the pure-state decomposition
theorem for density operators \cite{Schrodinger36,Jaynes57,Hughston93} 
says that a set of subnormalized vectors gives a decomposition of $\rho$ 
if and only if the vectors can be written as
\begin{equation}
\label{eq:trans}
|w_j\rangle=\sum_{k=1}^{n}O_{kj}|e_k\rangle,\;\;\;j=1,\ldots,m\;,
\end{equation}
where $O$ is an $n\times m$ matrix whose $n$ rows are real
$m$-dimensional orthonormal vectors.  We can always extend $O$ to be an
$m\times m$ orthogonal matrix by adding additional rows.

Now notice that the projector onto the support of $\rho$ can be written
as
\begin{equation}
\Pi=
\sum_{j=1}^m \rho^{-1/2}|w_j\rangle\langle w_j|\rho^{-1/2}\;,
\label{eq:Pi}
\end{equation}
where
\begin{equation}
\rho^{-1/2}|w_j\rangle=
\sum_{k=1}^{n}O_{kj}|\hat e_k\rangle,\;\;\;j=1,\ldots,m\;.
\end{equation}
By adding additional orthonormal vectors $|\hat e_k\rangle$ for
$k=n+1,\ldots,m$, we can define $m$ orthonormal vectors in an
extended Hilbert space,
\begin{equation}
|\overline w_j\rangle\equiv
\sum_{k=1}^{m}O_{kj}|\hat e_k\rangle,\;\;\;j=1,\ldots,m\;.
\end{equation}
Notice that
\begin{equation}
\Pi|\overline w_j\rangle=\rho^{-1/2}|w_j\rangle\;,
\end{equation}
which implies that
\begin{equation}
|w_j\rangle=\rho^{1/2}|\overline w_j\rangle\;.
\end{equation}
The extension of the resolution of $\Pi$ in Eq.~(\ref{eq:Pi}) to a
set of orthonormal vectors in a higher-dimensional space is called a
{\it Neumark extension} \cite{Peres95}.

Now we apply these concepts to a two-rebit state $\rho$ having the 
pure-state decomposition~(\ref{eq:psdecomp}).  The average
concurrence of this decomposition satisfies the inequality
\begin{equation}
\langle{C}\rangle=
\sum_j p_j C({\hat w}_j)=
\sum_j p_j|\langle \hat w_j|\sigma_y\otimes\sigma_y|\hat w_j\rangle|\ge
\biggl|\sum_j p_jc(\hat w_j)\biggr|=
|\langle c\rangle|
\;,
\end{equation}
where, following Wootters, we define the {\it preconcurrence\/} of the
pure
state $|\hat w_j\rangle$ without the absolute value signs that make the
concurrence positive:
\begin{equation}
c(\hat w_j)\equiv\langle\hat w_j|\sigma_y\otimes\sigma_y|\hat
w_j\rangle\;.
\end{equation}
The attractive feature of the preconcurrence is that its average value
is
independent of the ensemble decomposition, being given by
\begin{equation}
\langle c\rangle= \sum_j\langle
w_j|\sigma_y\otimes\sigma_y|w_j\rangle= \sum_j \langle\overline w_j|
\rho^{1/2}(\sigma_y\otimes\sigma_y)\rho^{1/2} |\overline w_j\rangle=
{\rm tr}(\tau) \;,
\end{equation}
where
\begin{equation}
\label{eq:tau}
\tau\equiv\rho^{1/2}(\sigma_y\otimes\sigma_y)\rho^{1/2}
\end{equation}
is a real, symmetric operator.

We now show that there is a pure-state ensemble whose average
concurrence
achieves the lower bound, $|{\rm tr}(\tau)|$.  We actually show
something
stronger, using the approach introduced by Wootters: there is a
pure-state
ensemble such that the preconcurrence of every member of the ensemble is
${\rm tr}(\tau)$ and, hence, the concurrence of every member of the
ensemble
is $|{\rm tr}(\tau)|$.  This stronger result becomes important when we
consider the entanglement of formation.  To construct the desired
ensemble,
start with the eigendecomposition of $\rho$.  If the eigendecomposition
has
only one member, then we are dealing with a pure state whose
preconcurrence
is ${\rm tr}(\tau)$, and nothing further needs to be done.  If the
eigendecomposition has more than one member, consider the states with
the largest and smallest preconcurrences.  Since the average
preconcurrence
is ${\rm tr}(\tau)$, the largest preconcurrence must be greater than or
equal to ${\rm tr}(\tau)$, and the smallest must be less than or equal
to ${\rm tr}(\tau)$.  Consider the continuous sequence of
two-dimensional
orthogonal matrices that range from the identity matrix to the matrix
that
exchanges the subnormalized vectors for the states with the largest and
smallest preconcurrences.  Somewhere along this sequence, the state with
the largest preconcurrence is transformed into one whose preconcurrence
is ${\rm tr}(\tau)$.  Adopt the resulting ensemble decomposition.  Now
iterate this procedure, always choosing the states with the largest and
smallest preconcurrences and transforming the state with the largest
preconcurrence to one with preconcurrence equal to ${\rm tr}(\tau)$.
The result is an ensemble decomposition all of whose members have
preconcurrence
equal to ${\rm tr}(\tau)$, as promised.

So far in this section we have shown that for a two-rebit density
operator
$\rho$, the minimum average concurrence over all the pure-state
ensembles
for $\rho$ is
\begin{equation}
\langle C\rangle_{\rm min}\equiv
\min_{\{p_j,|\hat w_j\rangle\}}\sum_jp_jC(\hat w_j)=
|{\rm tr}(\tau)|=
|{\rm tr}(\rho\sigma_y\otimes\sigma_y)|
\;.
\end{equation}
This justifies calling $|{\rm tr}(\tau)|$ the concurrence of $\rho$, as
in Eq.~(\ref{eq:realc}).

The entanglement of formation for a two-rebit state now follows with
very
little further work, since it satisfies the following chain of
relations:
\begin{equation}
E(\rho)\equiv
\min_{\{p_j,|\hat w_j\rangle\}}
\sum_j p_j{\cal E}[C(\hat w_j)]\ge
\min_{\{p_j,|\hat w_j\rangle\}}
{\cal E}\biggl(\sum_j p_j C(\hat w_j)\biggr)=
{\cal E}(\langle C\rangle_{\rm min})=
{\cal E}[|{\rm tr}(\tau)|]
\;.
\end{equation}
The inequality follows from the convexity of the function ${\cal
E}(C)$, and the immediately following equality follows from the
monotonicity of ${\cal E}(C)$.  To saturate the inequality requires
an ensemble all of whose members have a concurrence equal to
$\langle C\rangle_{\rm min}=|{\rm tr}(\tau)|$.  Having just
constructed an ensemble, we conclude that the entanglement of
formation of a two-rebit state $\rho$ is given by
\begin{equation}
E(\rho)={\cal E}(|{\rm tr}(\tau)|)={\cal E}[C(\rho)]\;.
\end{equation}

\section{Discussion}
\label{sec:disc}

To conclude, we reiterate that we now possess a complete expression
for the entanglement of formation of two rebits:
\begin{equation}
E(\rho^{\ho AB})=H\!\left(1+\sqrt{1-C^2(\rho^{\ho AB})}\over2\;\right)\;,
\end{equation}
where
\begin{equation}
C(\rho^{\ho AB})=|{\rm tr}(\tau)|
=|{\rm tr}(\rho^{\ho AB}\sigma_y\otimes\sigma_y)| \;.
\label{eq:realconcurrence}
\end{equation}
In particular, this expression implies that $\rho^{\ho AB}$ is
separable relative to real vector space if and only if 
${\rm tr}(\rho^{\ho AB}\sigma_y\otimes\sigma_y)=0$.
Notice that this separability condition is equivalent to saying that
$\rho^{\ho AB}$ is real separable if and only if it is unchanged by
partial transposition, i.e., 
\begin{equation}
\rho^{\ho AB}
= (\rho^{\ho AB})^{T_A} 
= (\rho^{\ho AB})^{T_B}\;,
\end{equation}
where $T_{\ho A}$ ($T_{\ho B}$) denotes partial transposition of system 
$A\,$($B$) in any orthonormal basis. 

It is worth stressing the difference between the 
expression~(\ref{eq:realconcurrence}) for the concurrence in a real
vector space and the Wootters formula~(\ref{eq:compconcurrence}) for
concurrence in standard quantum theory.  If we let $\nu_j$ be the
eigenvalues of $\tau$, ranked in order of decreasing absolute value,
the real concurrence is 
\begin{equation}
C(\rho^{\ho AB})=|\nu_1+\nu_2+\nu_3+\nu_4|\;.
\end{equation}
In contrast, for the Wootters concurrence, one first finds the eigenvalues 
of $\tau\tau^\dagger=\tau^2$, these being given by $\lambda_j=\nu_j^2$; then
the Wootters concurrence is 
\begin{equation}
C_W(\rho^{\ho AB})=\max(0,|\nu_1|-|\nu_2|-|\nu_3|-|\nu_4|)\le C(\rho^{\ho AB})\;.
\end{equation}

This difference is illustrated by the class of real states of the form 
\begin{equation}
\rho^{\ho AB}
=\frac{1}{4}\Big(I{\otimes}I+\alpha(\sigma_y\otimes\sigma_y)\Big)\;,
\label{eq:alphastates}
\end{equation}
where $\alpha$ is a positive real number that ranges from 0 to 1.  For these 
states, the operator $\tau$, given by 
\begin{equation}
\tau
=\frac{1}{4}\Big(\alpha(I{\otimes}I)+\sigma_y\otimes\sigma_y\Big)\;,
\end{equation}
has doubly degenerate eigenvalues ${1\over4}(\alpha\pm1)$.  The real 
concurrence is $C=\alpha$, whereas the Wootters concurrence is $C_W=0$.  
Thus these states are complex separable, but real entangled, except for 
$\alpha=0$.  Moreover, because these states are complex separable, they 
have no distillable entanglement, so their real entanglement is bound.

Our results can be considered a first step toward getting a better
understanding of which features of quantum entanglement are unique
to standard quantum mechanics and which are more generic across
various foil theories.  As just shown, the states~(\ref{eq:alphastates})
show that bound entanglement is sometimes nothing more than separability 
with respect to a larger field (in this case the complex numbers of standard 
quantum mechanics).  One might ask to what extent this is true of bound 
entanglement in standard quantum mechanics.  How many bound entangled states 
in standard quantum mechanics are bound because they are separable with 
respect to a quaternionic theory \cite{Adler95}?

Another interesting fact is how the regions of entangled vs.\ separable
states within the full set of quantum states differ in going from
real to complex quantum mechanics. In the complex theory, the maximally 
mixed state $\rho=\frac{1}{4} I_4$ of two qubits is surrounded by an open 
set of separable states \cite{Zyczkowski98,Braunstein99}.  In the real 
theory, however, the states~(\ref{eq:alphastates}) demonstrate that there 
are entangled states arbitrarily close to the maximally mixed state.  

Where all this will lead, we are not quite sure, but in general it
forms part of a larger effort to understand the nature of entanglement in 
our quantum world.

\acknowledgments We thank Patrick Hayden and Barbara Terhal for
useful discussions.  This work was supported in part by the
U.S.~Office of Naval Research (Grant No.~N00014-00-1-0578).

\end{document}